\documentclass[12pt]{article}
\usepackage{epsfig,amsfonts,amssymb}
\usepackage{hyperref}
\pdfoutput=1

\usepackage{cite}
\usepackage{cite}

\usepackage{comment}

\def\ZZZ{{\hbox{ Z\kern-1.6mm Z}}}
\def\RRR{{\hbox{ R\kern-2.4mm R}}}
\def\CCC{{\hbox{ C\kern-2.0mm C}}}
\def\zzz{{\hbox{z\kern-1mm z}}}

\def\ZZZ{\mathbb{Z}}
\def\RRR{\mathbb{R}}

\newcommand{\vt}{\vartheta}

\newcommand{\qeq}{{\hbox{=\kern-2.3mm ? \kern.5mm }}}
\renewcommand{\qeq}{=}

\newcommand{\eps}{\epsilon}

\newcommand{\BB}{{\cal B}}

\newcommand{\EE}{{\cal E}}

\newcommand{\XX}{{\cal X}}

\newcommand{\be}{\begin{equation}}
\newcommand{\ee}{\end{equation}}
\newcommand{\ben}{\begin{eqnarray}\displaystyle}
\newcommand{\een}{\end{eqnarray}}

\newcommand{\refb}[1]{(\ref{#1})}
\newcommand{\p}{\partial}
\newcommand{\sectiono}[1]{\section{#1}\setcounter{equation}{0}}

\def\one{{\hbox{ 1\kern-.8mm l}}}
\def\zero{{\hbox{ 0\kern-1.5mm 0}}}

\newcommand{\bea}[1]{\begin{eqnarray}\label{#1} }
\newcommand{\eea}{\end{eqnarray}}

\newcommand{\eqref}{\refb}

\usepackage{bm}
\usepackage[table]{xcolor}

\begin{document}

\baselineskip 24pt

\begin{center}

%{\Large \bf How to Study Isolated Exotic Branes}
{\Large \bf A New Term in Type II Effective Action}

\end{center}

\vskip .6cm
\medskip

\vspace*{4.0ex}

\baselineskip=18pt

\centerline{\large \rm Ashoke Sen}

\vspace*{4.0ex}

\centerline{\large \it International Centre for Theoretical Sciences - TIFR 
}
\centerline{\large \it  Bengaluru - 560089, India}

%\centerline{\large \it ~$^c$Homi Bhabha National Institute}
%\centerline{\large \it Training School Complex, Anushakti Nagar,
%    Mumbai 400085, India}

\vspace*{1.0ex}
\centerline{\small E-mail:  ashoke.sen@icts.res.in}

\vspace*{5.0ex}

\centerline{\bf Abstract} \bigskip

We show that the one loop 
effective action of type IIA and IIB string theories in ten dimensions have terms
proportional to the dilaton times the ten dimensional Euler density, in apparent violation
of the soft dilaton theorem. This can be traced to the existence of the zero modes of the world-sheet
superconformal ghost fields and resolves a puzzle
that arose recently in the analysis of black hole index when we compactify these theories on
Calabi-Yau manifolds of non-zero Euler number.

\vfill \eject

\tableofcontents

\section{Introduction} \label{s1}

In string theory, soft dilaton theorem tells us that if in a given amplitude we insert a zero
momentum dilaton vertex operator, we get back the same amplitude up to a constant of
proportionality\cite{9411047,9507038}. 
For insertion of the dilaton into a genus $g$, $n$-point amplitude of closed strings, 
the proportionality factor
is $(2g-2+n)$. Hence if we take $g=1$ and $n=0$ then the constant of proportionality vanishes
and the result vanishes. There is however a subtlety since for odd spin structure, i.e.\ when the
GSO odd world-sheet fields satisfy periodic boundary condition along both cycles,  
the torus partition function diverges due to zero modes of the bosonic $\beta$-$\gamma$ ghost 
system\cite{2408.14567,2505.14380,mslg}. Hence the
result for the dilaton one point function has the form $0\times \infty$ and becomes ambiguous.

The ambiguity, however, arises only when we try to apply soft dilaton theorem. We can instead
compute the dilaton one point function directly and this is neither divergent nor ambiguous.
We show that the genus one contribution to the dilaton one point function gets a non-vanishing
contribution proportional to the Euler number of the target space from  the odd-odd spin structure.
This can be summarized by saying that the ten dimensional effective action gets a contribution
of the form
\be \label{enew}
\mp {1\over 24}\, \int d^{10}x \, \phi(x)\, \EE(x)\, ,
\ee
where $\phi$ is the dilaton field, $\EE(x)$ is the ten dimensional Euler density 
and the top and bottom signs are valid for
type IIB / type IIA string theories. Our notation for the sign of the action is such that the
Euclidean path integral is directly weighted by the exponential of the action.

This resolves a puzzle that arose recently in the analysis of logarithmic correction to the 
BPS entropy of a black hole in type II string theory on Calabi-Yau three fold, carrying purely 
RR charges\cite{2606.17149}.
The string coupling is an arbitrary parameter in this solution and hence the BPS entropy, being the
logarithm of an index, must be independent of the string coupling. Explicit computation on the other
hand gave a result
\be\label{etwo}
\pm {1\over 12} \, \chi_{CY}\, \ln g_s\, ,
\ee
where $\chi_{CY}$ is the Euler number of the Calabi-Yau manifold. We now see that the
contribution \refb{enew} is precisely what is needed to cancel this term since the Euclidean
black hole space-time has Euler number 2 and hence $\int d^{10}x\, \EE(x)=2\, \chi_{CY}$ for
these solutions..

We note that \refb{enew} is not the only way that the dilaton dependence can enter an amplitude.
Typically terms proportional to $\phi$, i.e.\ $\ln g_s$, arise from infrared effects, when we express
logarithm of the energy measured in string scale in terms of logarithm of energy measured in the
Planck scale\cite{1002.3805,1810.11343}. 
These terms are very similar in spirit to the ones computed in \cite{2606.17149} and
summarized in \refb{etwo}.  The analysis of the current paper deals with a different type of 
$\ln g_s$ term
that arises even when we express the result in terms of the lengths and energies measured in
string scale. Indeed the insertion of a zero momentum dilaton vertex operator in a correlation function
computes the derivative of the amplitude without the dilaton with respect to $\ln g_s$ for fixed background
{\it string frame metric} and fixed momenta of external states {\it measured in string scale}.
As already mentioned, for black hole solutions in these string theories compactified on
Calabi-Yau three folds, 
the kind of logarithmic terms found in this paper seem to be
needed to cancel some unwanted effects of the  terms coming from the infrared effect.

Before turning to a description of the organization of the paper, we would like to add a word of caution. 
As will be discussed in the text, there
are two sign issues, -- in eqs.\refb{e6} and \refb{e212}, --  that we have not been able to fix to our complete
satisfaction. Nevertheless, we expect \refb{etwo} to be correct since this is what is needed to cancel
the unwanted term in the black hole index. Without this cancellation the index will depend on the
asymptotic value of the string coupling, which is inconsistent with space-time supersymmetry.

The rest of the paper is organized as follows. In section \ref{s2} we set up the computation of the 
dilaton one point function on the torus and show that naively the result vanishes by ghost
number conservation. We also identify the source of an error in this argument as due to the result
of vertical integration\cite{1408.0571,1504.00609} that is needed to maintain modular invariance. 
In section \ref{s3} we evaluate the contribution due to the vertical integration and
arrive at \refb{enew}. 
The analysis of this section is restricted to the case where the spin structures are
chosen to be odd both in the holomorphic and the anti-holomorphic sectors.
In section \ref{s4} we show that the contributions to the dilaton one point function from all other spin
structures vanish.

It will be interesting to explore if the $\ln g_s$ terms found here are related to the ones in topological string
partition function\cite{9302103} in any way.

\sectiono{Bulk integral} \label{s2}

We shall begin by reviewing various normalization conventions. We shall follow the conventions of
\cite{2405.19421} and hence will write down only those results that we shall use repeatedly.
 The picture changing operators will be taken to be
 \be \label{e3}
\XX=  c\p\xi + e^\phi T_F -{1\over 4}\, \p\eta e^{2\phi} b -
{1\over 4}\, \p\bigl( \eta e^{2\phi} b\bigr), \qquad 
\bar\XX =  \bar c\bar\p\bar\xi + e^{\bar\phi} \bar T_F -{1\over 4} \bar\p \bar\eta e^{2\bar\phi} \bar b -
{1\over 4}\bar\p\bigl( \bar\eta e^{2\bar\phi} b\bigr)\, .
\ee
Here $\xi,\eta$ are the grassmann odd fields and $\phi$ is the grassmann even field arising from the
bosonization of the $\beta$, $\gamma$ superghosts. $\bar\xi,\bar\eta,\bar\phi$ are their
anti-holomorphic counterparts. $b,c,\bar b,\bar c$ are the usual grassmann odd ghost fields
arising from world-sheet diffeomorphism invariance.

Next we shall construct the zero momentum dilaton vertex operator.
In superstring theory this is known to be given by (see {\it e.g.} \cite{2506.00120})
\be \label{e6}
V_D=-4\, \left(\eta c \, \bar c \bar\p\bar \xi e^{-2\bar\phi} - \bar \eta \bar c \, c \p\xi e^{-2\phi} 
\right)\, .
\ee
Ref.~\cite{2506.00120} did not have the factor of 4 since they use a different normalization convention, but
this can be fixed using the translation rule given in footnote 3 of that paper, which effectively
amounts to
shifting $\phi$ and $\bar\phi$ by $-\ln 2$. 
The first sign issue mentioned in the introduction arises
here. If we apply
the picture changing operator
$\XX\bar\XX$ to convert this to a zero picture vertex operator, we get\footnote{The singular parts appearing during
the picture changing operation can be removed by averaging over symmetric placement of the
picture changing operator, e.g. $\XX(0)V(0)$ is interpreted as $\ointop dw w^{-1} \XX(w) V(0)$ where $w$
is the local coordinate in which the vertex operator $V$ is inserted and $\ointop$ denotes contour integral around
$w=0$, normalized so that $\ointop dw w^{-1}=1$.}
\be \label{e4}
-{1\over 2}\,  (c\p^2 c - \bar c \bar\p^2 \bar c) +\cdots\, ,
\ee
where $\cdots$ denotes other terms containing powers of $e^\phi$.
On the other hand, 
according to \cite{2405.19421}, sections 4.7 and 5.7, the vertex operator of zero momentum 
dilaton in boosnic string theory is given by
\be\label{e1}
{1\over 2} (c\p^2 c - \bar c \bar\p^2 \bar c)\, .
\ee
This differs from \refb{e4} by a sign. We shall proceed with \refb{e6} since it is determined by comparing the
string field theory action with the expected form of the low energy effective action and does not involve any
comparison with the bosonic string theory.

Our goal will be to compute the dilaton one point function on the torus. We shall label the torus by
the coordinate $w$ with the identification:
\be \label{e7}
w\equiv w+1 \equiv w+\tau\, .
\ee
The coordinate system $z$ in which the torus can be seen as a propagating closed string of length $2\pi$
is given by
\be 
z = 2\pi w \, .
\ee
In the $z$ coordinate system $b(z)$ and $c(z)$ has the expansion
\be
b(z) =\sum_n b_n \, e^{ in z}, \qquad c(z) =\sum_n c_n \, e^{i nz}\, .
\ee
Converting this to the $w$ coordinate system we get
\be\label{e210}
b(w) = b(z) (dz/dw)^2 = 4\pi^2 \sum_n b_n \, e^{2\pi i n w}, \qquad
c(w) = c(z) (dz/dw)^{-1} = {1\over 2\pi} \sum_n  c_n \, e^{2\pi i n w}\,.
\ee
Similarly we get
\be\label{e211}
\bar b(\bar w) = 4\pi^2 \sum_n \bar b_n \, e^{-2\pi i n \bar w}, \qquad
\bar c(\bar w) =  {1\over 2\pi} \sum_n  \bar c_n \, e^{-2\pi i n \bar w}\,.
\ee

Next we fix the normalization of the torus partition function. 
We shall work in the sector where the spin structure is odd in both the holomorphic and the
anti-holomorphic sectors. This means that all GSO odd world-sheet fields have periodic boundary
condition along all the cycles of the torus.
We shall use the convention that in
the large Hilbert space the correlation functions are computed by inserting a factor of $\xi(x) \bar\xi(y)$
to the extreme left of the small Hilbert space 
correlator. With this convention the relevant torus
correlation function in the large Hilbert space takes the form
\ben \label{e212}
&& \left\langle \xi(x_1) \xi(x_2) \eta(y_1) e^{q\phi}(z_1) e^{-q\phi}(z_2) \ 
\bar \xi(\hat x_1) \bar\xi(\hat x_2) \bar\eta(\hat y_1) e^{p\bar\phi}(\hat z_1) e^{-p\bar\phi}(\hat z_2)
b_0c_0\bar b_0 \bar c_0 
\right\rangle_L \nonumber \\
&=& \mp\chi\, \eta(\tau)^3 (\eta(\tau)^*)^3 \times
 \left\langle \xi(x_1) \xi(x_2) \eta(y_1) e^{q\phi}(z_1) e^{-q\phi}(z_2)
\right\rangle_{\rm hol} \nonumber \\ &\times&
   \left\langle\bar \xi(\hat x_1) \bar\xi(\hat x_2) \bar\eta(\hat y_1) e^{p\bar\phi}(\hat z_1) e^{-p\bar\phi}(\hat z_2) \right\rangle_{\rm anti-hol}\, .
   \een
Here $\chi$ is the Euler number of the target space,    
arising due the fact that in the 
matter sector the partition function in the
odd spin structure is given by the Witten index which is the Euler number of the target space.
As will be explained shortly, 
in \refb{e212}  the top sign is for type IIB string theory where we have
identical GSO projection rules in the holomorphic and the anti-holomorphic sector, while the
bottom sign is for type IIA theory where we have opposite GSO projection rules in the
holomorphic and the anti-holomorphic sectors.
For later use, we shall write down the holomorphic and the anti-holomorphic correlation functions
for general spin structure $\delta$\cite{verlinde}:
\ben\label{e213}
&& \left\langle \xi(x_1) \xi(x_2) \eta(y_1) e^{q\phi}(z_1) e^{-q\phi}(z_2)
\right\rangle_{\rm hol}\nonumber \\
 &=& {\vt_1(x_1-x_2)  \vt_1(z_1-z_2)^{q^2} \over \vt_1(x_1-y_1) \vt_1(x_2-y_1)}
\, \vt_1'(0)^{1-q^2}\, {\vt_\delta(x_1+x_2-2y_1 + q z_1-q z_2)\over \vt_\delta(x_1-y_1 + q z_1-q z_2)
\vt_\delta(x_2-y_1 + q z_1-q z_2)} \, . \nonumber \\
\een
$\langle ~\rangle_{\rm anti-hol}$ is given by the complex conjugate of the holomorphic
correlation function.
Here $\vt_\delta$ are the Jacobi theta functions
with $\vt_1$ being the odd
Jacobi theta function,
$\eta(\tau)$ is the Dedekind function and
$*$ denotes complex conjugation.\footnote{We shall normalize the arguments of the $\vt_i$'s such that
they are quasi-periodic in $z$ with periods 1 and $\tau$, and $\vt_1'(0)=2\pi \eta(\tau)^3$ where $'$ denotes
derivative with respect to $z$. The overall phase in \refb{e213} will cancel between the holomorphic and the
anti-holomorphic sector and will not change the final result.}
In subsequent calculations in this and the next section we shall set $\delta=1$ since we
are working in the odd spin structure sector. 
The $\eta(\tau)^3$ is the contribution from the non-zero mode 
oscillators of the holomorphic  fields. There are four
grassmann odd  fields $b$, $c$, $\xi$, $\eta$ and one grassmann even
field $\phi$, explaining the power 3.  The rest of the factors were found in \cite{verlinde}.

Let us now examine the sign of \refb{e212} carefully. Before GSO projection the RR sector ground states in the
matter sector are in the
\be
(16\oplus\overline{16}) \otimes (16\oplus\overline{16})
\ee
representation of the tangent space $SO(10)$ group of the Euclidean space-time. Using the
representation theory of $SO(10)$ one finds that the $16\otimes 16$ and 
$\overline{16}\otimes \overline{16}$ representations give the odd rank anti-symmetric tensors of
$SO(10)$ while $16\otimes \overline{16}$ and $\overline{16}\otimes 16$ representations give
the even rank anti-symmetric tensors of
$SO(10)$. If we denote by $F$ the total world-sheet fermion number including the left and the
right chiral sectors, then in type IIB string theory the former have $(-1)^F$ eigenvalue 1 while the latter have 
$(-1)^F$ eigenvalue $-1$. Hence $Tr(-1)^F$ gives $-\chi$. In type IIA string theory the $(-1)^F$
eigenvalues are opposite and hence $Tr(-1)^F$ gives $\chi$.

Note, however, that this analysis relies on counting states in the $(-1/2,-1/2)$ picture where we
naturally get gauge field strength instead of the gauge potential. In $(-3/2,-1/2)$ or $(-1/2,-3/2)$ picture
the counting will be different and we shall get gauge fields instead of gauge field strength. This will
produce opposite sign in \refb{e212}. This is the second sign issue mentioned in the introduction.
We have used the $(-1/2,-1/2)$ picture states since we shall explicitly insert a pair of PCOs 
given in \refb{e216}, which have non-zero matrix element between a pair of states in the 
$(-3/2,-3/2)$ picture, which, in turn, are conjugate to $(-1/2,-1/2)$ picture states.

We can use \refb{e213} to show that the partition function diverges. For this we can set $q=0$ and
take the limit $x_2\to y_1$ after multiplying the right hand side by $(x_2-y_1)$ so that using
operator product expansion we can remove the $\xi(x_2)\eta(y_1)$ factor and are left with 
just $\xi(x_1)$ to absorb the $\xi$ zero mode. However since the right hand side of \refb{e213} has
a double pole at $x_2=y_1$, this expression diverges. This can be traced to the existence of the
$\beta$-$\gamma$ zero modes in the odd spin structure.

We can use \refb{e213} to find the small Hilbert space 
correlation function $\langle\cdots \rangle_S$ of $\p\xi$, $\eta$ and $e^{\pm q\phi}$.
For this we take the derivative of \refb{e213} with respect to $x_2$. Since the result is expected to be
independent of $x_1$, we can simplify the analysis by setting $x_1=x_2$. This gives:
\ben \label{e214}
&& \left\langle \p\xi(x_2)  \eta(y_1) e^{q\phi}(z_1) e^{-q\phi}(z_2)\right\rangle_{\rm hol;S}
\nonumber \\
&=& - {\vt_1(z_1-z_2)^{q^2} \over \vt_1(x_2-y_1)^2}
\times \vt_1'(0)^{2-q^2}
\times {\vt_1(2x_2-2y_1 + q z_1-q z_2)\over 
\vt_1(x_2-y_1 + q z_1-q z_2)^2}\, . \nonumber \\
\een
As is customary, we have dropped the $\xi(x_1)$ factor on the left hand side 
since the result is independent of $x_1$.

In order to compute the dilaton one point function, we need to integrate an appropriate correlation
function over the fundamental region of the moduli space of the torus, labelled by the complex parameter
$\tau=\tau_1+i\tau_2$ satisfying
\be \label{e215}
-{1\over 2}\le\tau_1<{1\over 2}, \qquad |\tau|\ge 1\, .
\ee
Besides the $(-1,-1)$ picture dilaton vertex operator, we need to insert some more operators 
into the correlator. First we need to insert two picture changing operators, one in the holomorphic
sector and one in the anti-holomorphic sector. Since we are working in the odd spin structure sector,
in the $\tau\to i\infty$ limit, that is included in the fundamental region \refb{e215}, we have RR sector
states propagating along the long tube along the $b$-cycle. In this limit the PCO insertions must be averaged over the
$a$-cycle of the torus in order to obtain the correct Ramond sector
propagator\cite{1209.5461,1501.00988,1508.05387}:
\be \label{e216}
\int_{c_1}^{c_1+1} dy_1 \ \XX(y_1)  \int_{c_2}^{c_2+1} dy_2 \ \bar\XX(y_2)\, ,
\ee
where $c_1$ and $c_2$ are arbitrary points on the torus. We choose them so that the contours
do not pass through the dilaton vertex operator. Away from $\tau\to i\infty$ limit we can
choose different PCO locations, but we shall keep them as in \refb{e216} to simplify the
analysis. We shall see later that this will lead to an error that will have to be rectified
by vertical integration.

The second set of insertions involve $b$-ghosts to produce the correct integration measure.
Using the results reviewed in \cite{2405.19421} and the identification \refb{e7} we see that the
required insertion is
\be
d\tau\wedge d\bar\tau\, \BB_\tau \BB_{\bar\tau}\, ,
\ee
where
\be
\BB_\tau = -{1\over 2\pi i} 
\int_0^1 dw b(w)=2\pi i  b_0, \qquad \BB_{\bar\tau} = {1\over 2\pi i} \int_0^1 d\bar w \, \bar b(\bar w)
=-2\pi i \bar b_0\, .
\ee
where in the last steps we have used \refb{e210}, \refb{e211}. In writing this we have taken into account
the fact that the coordinate on the left of the $w$ integration contour is equal to 
the coordinate on the right of the $w$ integration contour minus $\tau$.

Finally we have to determine the overall normalization of the amplitude. First of all, one can
use the prescription reviewed in \cite{2405.19421} to conclude that the 
genus one one point function will have
a factor of $(-2\pi i)^{-1}$. Second, since for GSO projection we need to average over periodic
and anti-periodic boundary conditions along the $b$-cycle, we get a factor of 1/2 each from the
holomorphic and the anti-holomorphic sectors. Finally due to the $w\to -w$ symmetry of the
torus one point function (assuming that the vertex operator is inserted at $w=0$) we get another
factor of $1/2$. 

Combining these results together, we arrive at the result:
\be\label{e219}
-{1\over 2\pi i} \times {1\over 2} \times {1\over 4} \times \int d\tau \wedge d\bar \tau 
\left\langle \xi(x_1)\bar\xi(\hat x_1)  \BB_\tau \ \BB_{\bar\tau} \ 
\int_{c_1}^{c_1+1} dy_1 \ \XX(y_1)  \int_{c_2}^{c_2+1} dy_2 \ \bar\XX(y_2) \ V_D\right\rangle\, .
\ee
It is easy to see that the integrand vanishes due to ghost number non-conservation. 
Indeed $V_D$ given in \refb{e6} has two terms, one
with ghost number (0,2) and the other with ghost number (2,0), whereas the other operators 
inserted into the correlator have combined ghost number $(-1,-1)$ from the $\BB_\tau\BB_{\bar\tau}$
factor.\footnote{$c$ and $\eta$ have ghost number (0,1) and $b$ and $\xi$ have ghost number 
$(0,-1)$. Similar assignments hold in the anti-holomorphic sector. 
We ignore the $\xi(x_1)\bar\xi(\hat x_1)$ factor in \refb{e219}
from this counting; they would be
absent in the small Hilbert space representation anyway.}

If this was the full story, then we would conclude that the dilaton one point function on the
torus vanishes. However, there is a subtlety in this analysis that arises from the assumption
that the PCO insertions can be taken to be of the form \refb{e216} everywhere in the moduli
space. The reason is that
in the fundamental region \refb{e215}, the modular transformation $\tau\to -1/\tau$ 
exchanges the $\tau_1\ge 0$ and $\tau_1\le 0$ segments of the $|\tau|=1$ boundary.
If we parametrize $\tau$ by
\be
\tau=r e^{i\theta}\, ,
\ee
then the two boundary segments correspond to $r=1$, $\pi/3\le \theta\le \pi/2$ and $\pi/2\le\theta\le 2\pi/3$
respectively, and the $\tau\to -1/\tau$ symmetry exchanges these two segments by
identifying $\theta$ with $\pi-\theta$. Now
since $\tau\to -1/\tau$ transformation exchanges the $a$ and the $b$ cycles, we see that it is
inconsistent to take the PCO insertions to be averaged over the $a$-cycle on both segments
of the boundary, --  the integrand is not invariant under $\tau\to -1/\tau$ with this prescription.

In the next section we shall resolve this problem by carrying out a 
vertical integration that makes the PCO
insertion jump from average over the $a$-cycle to average over the $b$-cycle across the
boundary segment
\be
r=1, \qquad \pi/2\le\theta\le 2\pi/3\, .
\ee
We shall see that this produces a non-zero contribution to the dilaton one point function.

\sectiono{Vertical integration} \label{s3}

We begin by expressing the bulk integration measure as
\be
d\tau \wedge d\bar \tau \, \BB_\tau \ \BB_{\bar\tau}
= dr\wedge d\theta \, \BB_r \, \BB_\theta
\ee
where
\be
\BB_r = {\p\tau\over \p r} \BB_\tau + {\p\bar\tau\over \p r} \BB_{\bar\tau}
= r^{-1} (\tau\BB_\tau + \bar\tau\BB_{\bar \tau}), 
\qquad 
\BB_\theta = {\p\tau\over \p \theta} \BB_\tau + {\p\bar\tau\over \p \theta} \BB_{\bar\tau}
= i\, (\tau\BB_\tau - \bar\tau\BB_{\bar \tau})\, .
\ee
Since the vertical integration will be carried out across the boundary $r=1$, following the
usual rules of vertical integration we have to drop the $\BB_r$ factor and replace one of the
PCOs (say $\XX$) by $(\xi(y_i)-\xi(y_f))$ where $y_i$ and $y_f$ are the initial and the final
positions of the PCO\cite{1408.0571,1504.00609}. 
This has to be done in turn for both the PCOs, and the final result can be
shown to be independent of the order in which we move them. If we take the initial PCO position to be the
average along the $a$-cycle and the final PCO position to be the average along the $b$-cycle, then 
there is an extra minus sign since the jump occurs as we move from larger value of $r$ to smaller
value of $r$. Therefore the movement of the location of $\XX$ and $\bar\XX$ will produce,
respectively, factors of
\be \label{e33}
\int_0^1 ds \left[  \xi\left(s\tau+c_3\right) - \xi\left(s+c_1\right)
\right], \qquad  \int_0^1 ds \left[  \bar\xi\left(s\tau+c_4\right) - 
\bar\xi\left(s+c_2\right)
\right]\, ,
\ee
where $c_i$'s are arbitrary complex constants chosen so that the contours do not pass through
the location of the vertex operator.
Using $\xi$-$\eta$ ghost number conservation, we can easily see that the first term will contribute
only when we pick the first term in the expression for the dilaton vertex operator \refb{e6}. Similarly
the second term will contribute only when we pick the second term in \refb{e6}.

We first focus on the first term in \refb{e6}. Hence we pick the first term in \refb{e33}.
It is easy to see using the $b$-$c$ ghost number
conservation that we need to
pick the term $i\tau\BB_\tau=-2\pi \tau b_0$  from $\BB_\theta$. Using \refb{e219}, \refb{e212}, \refb{e6}
this leads to the
result:
\ben\label{e34}
I&\equiv&   - {1\over 16\pi i} \times  \int d\theta 
\bigg\langle \xi(x_1)\bar\xi(\hat x_1) \,  \int_0^1 ds \left[  \xi\left(s\tau+c_3\right) - 
\xi\left(s+c_1\right)
\right]\, (-2\pi \tau b_0) 
\nonumber \\ &&
  \int_{c_2}^{c_2+1} dy_2 \ \bar\XX(y_2) \, (-4)  \, \eta c \, \bar c \bar\p\bar \xi e^{-2\bar\phi}(y_1)\bigg\rangle
\nonumber \\
  &=&   -
  {1\over 8 i} \times \int d\theta \, \tau\, 
\bigg\langle \xi(x_1)\bar\xi(\hat x_1) \,   b_0 \, \int_0^1 ds \left[  \xi\left(s\tau+c_3\right) - 
\xi\left(s+c_1\right)
\right]\, \nonumber \\ && 
  \int_{c_2}^{c_2+1} dy_2 \ \bar\p\bar \eta \bar b \, e^{2\bar\phi}(y_2)  \, \eta c
   \bar c \bar\p\bar \xi e^{-2\bar\phi}(y_1)\bigg\rangle \nonumber \\
  &=&  -
  {1\over 8 i} \times \int d\theta \, \tau\, 
\bigg\langle \xi(x_1)\bar\xi(\hat x_1) \,   b_0 c_0 \bar b_0 \bar c_0 
\, \int_0^1 ds \left[  \xi\left(s\tau+c_3\right) - 
\xi\left(s+c_1\right)
\right] \eta(y_1)\,  \nonumber \\ &&
  \int_{c_2}^{c_2+1} dy_2 \ \bar\p\bar \eta  e^{2\bar\phi}(y_2) 
    \bar\p\bar \xi e^{-2\bar\phi}(y_1)\bigg\rangle
 \nonumber \\
 &=& \pm {1\over 8i} \, \chi \,  \int_{\pi/2}^{2\pi/3} d\theta \, \tau \,
 \eta(\tau)^3 (\eta(\tau)^*)^3 \, (I_2 - I_1)\, I_3^*\, ,
\een
where $*$ denotes complex conjugation and
\be
I_1\equiv \int_{c_1}^{1 + c_1} dz \, \bigg\langle \xi(x_1) \xi(z) \eta(y_1)\rangle_{\rm hol}
= \int_{c_1}^{1 + c_1} dz \,  {\vt_1(x_1- z) \vt_1(x_1+z-2y_1) \vt_1'(0)\over \vt_1(x_1-y_1) \vt_1(z-y_1)
\vt_1(x_1-y_1) \vt_1(z-y_1)}\, ,
\ee
\ben
I_2 &\equiv&  \tau^{-1} \, \int_{c_3}^{\tau + c_3} dz \, \bigg\langle \xi(x_1) \xi(z) \eta(y_1)\rangle_{\rm hol}
\nonumber \\
&=& \tau^{-1}\int_{c_3}^{\tau+c_3} dz \,  {\vt_1(x_1- z) \vt_1(x_1+z-2y_1) \vt_1'(0)\over \vt_1(x_1-y_1) \vt_1(z-y_1)
\vt_1(x_1-y_1) \vt_1(z-y_1)}\, ,
\een
and, using \refb{e214},
\ben
I_3 &=& \int_{c_2}^{c_2+1} dy_2 \left\langle\  \bar\xi(\hat x_1) \bar\p\bar \eta  e^{2\bar\phi}(y_2) 
   \,   \bar\p\bar \xi e^{-2\bar\phi}(y_1)\right\rangle_{\rm anti-hol}\nonumber \\
&=& \vt_1'(0)^{-2}
\int_{c_2}^{c_2+1} dy_2 \left[{\p\over \p y_2} \left\{{\vt_1(y_2'-y_1)^4 \over \vt_1(y_1-y_2)^2}
\times {\vt_1(2y_2'-2y_2)\over \vt_1(2 y_2'-y_1 - y_2)^2  }\right\}
\right]_{y_2'=y_2}\, .
\een
Evaluating these integrals one finds
\be
I_2- I_1 =  {2\pi i\over \tau} {1\over \vt_1'(0)}={i\over \tau} {1\over \eta(\tau)^3} \, ,
\ee
and
\be 
I_3 = -2 \,  \vt_1'(0)^{-1} = -{1\over \pi \, \eta(\tau)^3}\, .
\ee
In particular, for $I_3$, the integrand becomes $y_2$ independent. Hence even if the $y_2$
integral averaged over the $b$-cycle, the answer will remain the same. This shows that
if we had first moved the PCO $\bar\XX$ to the $b$-cycle and then moved the PCO $\XX$, we
would get the same result.

Substituting these results into \refb{e34} we get
\be
I= \mp {\chi\over 8\pi} \int_{\pi/2}^{2\pi/3} d\theta=\mp {\chi\over 48}\, .
\ee

We also have to evaluate the contribution where we pick the second term in \refb{e6}.
This time the contribution comes from the vertical integration over the location of the anti-holomorphic
PCO, since the jump in the holomorphic PCO location will produce a factor involving the difference of $\xi$ along
the two contours, and due to the presence of $\p\xi$ term in the second term 
of the dilaton vertex operator the resulting correlator
will vanish by $\xi$-$\eta$ ghost conservation law.
The analysis involving the term where the anti-holomorphic PCO jumps is identical to the case we have 
analyzed, with the roles of holomorphic and anti-holomorphic terms exchanged. 
At the end it gives an identical contribution. Thus the final result is
\be
\mp {\chi\over 24}\, .
\ee

This establishes that the one loop correction to the effective action of type II string theory
includes a term
\be 
\mp {1\over 24} \int d^{10}x \, \phi\, \EE\, ,
\ee
where $\EE$ is the ten dimensional Euler density whose integral gives the Euler number of the
ten dimensional manifold.

\sectiono{Other spin structures} \label{s4}

In this section we shall analyze the contribution from the other spin structures and argue that their
contribution to the dilaton one point function vanishes. For this we shall restrict to the first term in the
dilaton vertex operator \refb{e6}, the analysis of the second term would involve exchanging the roles of
holomorphic and anti-holomorphic variables.

First we note, following the same argument that was given in section \ref{s3},
that as long as the PCO integration contour is kept independent of the moduli, the ghost
number conservation makes the integrand vanish. Therefore we need vertical integration to get a
non-zero contribution. Next
we note, again following the analysis in section \ref{s3}, that if the vertical integration occurs for the
anti-holomorphic PCO, then the resulting small Hilbert space correlation function will have two 
$\bar\xi$'s and no $\bar\eta$ and the correlator will vanish by the $\bar\xi$-$\bar\eta$ ghost
number conservation. So the vertical integration has to take place over the holomorphic PCO
location.

We first examine the holomorphic side of the correlation function. 
First suppose that in the holomorphic sector we have even spin structure. In this case modular transformations
do not keep the spin structure invariant but transform the three even spin structures into each other. We shall
start with the spin structure where we have periodic boundary condition along the $a$-cycle and anti-periodic
boundary condition along the $b$-cycle and denote this by $P_aA_b$. Since for large $\tau_2$ we have a
Ramond sector state propagating along the long throat, the PCO insertion has to be an integral  over the
$a$-cycle. For definiteness let us place the dilaton vertex operator at 0 and take the PCO integration contour
to pass through the point $\tau/2$. 
We express this as
\be\label{e41}
P_aA_b: \qquad z={\tau\over 2} +  s, \qquad 0\le s < 1\, .
\ee
This spin structure and the integration contour remains 
unchanged under $\tau\to\tau+1$. Hence there is no vertical integration needed at those boundaries.
However, across the boundary $|\tau|=1$ we encounter a modular transformation 
$\tau\to -1/\tau$,
$z\to z/\tau$, that transforms \refb{e41}  to
\be\label{e42}
A_aP_b: \qquad z={1\over 2} + \,\tau\,  s\equiv f_1(s), \qquad 0\le s < 1\, .
\ee
We can compare this with the desired PCO location. Since we have anti-periodic boundary
condition along the $a$-cycle, we are now in the NS sector and in the $\tau\to i\infty$ limit we need
to keep the PCO in the vicinity of the dilaton vertex operator. We take this to be the choice of 
PCO location in
the whole of the fundamental region. Let us denote the PCO location as
\be
z = f_2(s), \qquad 0\le s<1\, .
\ee
Then across the boundary $|\tau|=1$ we have a vertical jump in the PCO
location from $z=f_1(s)$ to $z=f_2(s)$ and the relevant part of the holomorphic correlation function
is proportional to
\be 
\int_0^1 \, ds\,\langle \xi(x_1) (\xi(f_2(s))-\xi(f_1(s))\eta(0)\rangle_{hol}\, .
\ee
Since the correlator is expected to be independent of $x_1$,  we can take $x_1=f_1(s)$ and express this as
\be \label{e45}
\int_0^1 \, ds\,\langle \xi(f_1(s))\, \xi(f_2(s))\eta(0)\rangle_{hol}
= \int_0^1 ds {\vt_1(f_1(s)-f_2(s))\over \vt_1(f_1(s))\vt_1(f_2(s))} \times {\vt_\delta(f_1(s)+f_2(s))\over \vt_\delta(f_1(s))
\vt_\delta(f_2(s))}\, ,
\ee
where we used \refb{e213}.
Now we note from \refb{e42} that
\be
f_1(1-s) = -f_1(s)+1+\tau\equiv - f_1(s)\, ,
\ee
with the equivalence sign signifying that the correlator is manifestly invariant under a shift of $f_1(s)$ by
1 and $\tau$. On the other hand $f_2(s)$, whose choice has been left unspecified so far, can be chosen
to be odd under $s\to 1-s$, e.g. we can choose
\be\label{e47}
f_2(s)=\cases{\eps \quad \hbox{for} \quad 0\le s\le {1\over 2} \cr
-\eps \quad \hbox{for} \quad {1\over 2}\le s \le 1\, .
}
\ee
With this choice the integrand in \refb{e45} becomes an odd function under $s\to 1-s$ when $\delta$ represents
an even spin structure and hence the
integral vanishes.

Finally note that the $A_aA_b$ spin structure is related to the $A_aP_b$ spin structure by a $\tau\to \tau-1$ modular
transformation across the boundary $\tau=1$. However since in the $\tau\to i\infty$ limit with $A_aA_b$ spin
structure we still have an NS sector state propagating along the long throat, we can choose the PCO location as
in \refb{e47}, and hence there is no vertical jump across this boundary. As in the case of $P_aP_b$ spin structure,
the two segments of the $|\tau|=1$ boundary are now identified by the transformation $\tau\to -1/\tau$, $z\to z/\tau$.
Under this the pair of points $\pm \eps$ transform to $\pm\eps/\tau$ and hence we need to make a vertical
jump from $\pm\eps$ to $\pm\eps/\tau$ across one of the segments. This contribution vanishes by the same
symmetry argument that led to the vanishing of \refb{e45}.

This shows that in order to get a non-zero contribution, 
we need to choose odd spin structure in the holomorphic sector. In that case the analysis
of vertical integration takes the same form as in section \ref{s3}. Since the (odd,odd) spin structure was already
analyzed in section \ref{s2} what remains is to analyze the contribution when the anti-holomorphic spin structure
is even. In this case the relevant part of the $\xi,\eta,\phi$ 
correlation function in the anti-holomorphic sector is proportional to
\be
\langle \bar\xi(\hat x_1) \bar\p\bar\eta e^{2\bar\phi} (y_2) \bar\p\bar\xi e^{-2\bar\phi}(0)\rangle_{anti-hol}\, .
\ee
Using the generalization of \refb{e213} to the anti-holomorphic sector,
we can see that this function is odd under $y_2\to -y_2$ for even spin structure in the
anti-holomorphic sector. 
Therefore, if we choose the 
PCO averaging procedure such that the integration range is invariant under $y_2\to -y_2$ then the result
vanishes. Such a choice is clearly possible, {\it e.g.} we make the choice \refb{e41} in the $P_aA_b$ sector
and \refb{e47} in the $A_aP_b$ and the $A_aA_b$ sector. This shows that the contribution vanishes even when
the anti-holomorphic sector has even spin structure.

This shows that the only non-vanishing contribution to the dilaton one point function comes from the sector
where we have odd spin structure both in the holomorphic and the anti-holomorphic sector.

\bigskip

\noindent{\bf Acknowledgement:} 
I wish to thank 
Sergey Alexandrov, Carlos Mafra, Raji Mamade, Jan Manschot, Boris Pioline,  
Mukund Rangamani, Edward Witten and Barton Zwiebach for useful discussions.
This work was supported by the ICTS-Infosys Madhava 
Chair Professorship and
Department of Atomic Energy, Government of India, under project no.~RTI4019.
I also acknowledge the hospitality of the Strings 2026 conference, during which part of this work was done.

\end{document}